\documentclass[]{tPHM2e}

\begin{document}
\doi{10.1080/14786435.20xx.xxxxxx}
\issn{1478-6443}
\issnp{1478-6435}
\jvol{00} \jnum{00} \jyear{2010} %\jmonth{21 December}

\def\onlinecite{\cite}
\def\tr{\textrm{tr}}
\def\bea{\begin{eqnarray}}
\def\eea{\end{eqnarray}}
\def\beq{\begin{equation}}
\def\eeq{\end{equation}}
\def\openone{\mathbb{I}}
\def\t{t}

\markboth{Taylor \& Francis and I.T. Consultant}{Philosophical Magazine}

%\articletype{GUIDE}

\title{{\itshape Philosophical Magazine} -- Topological Quantum Glassiness}

\author{Claudio Castelnovo$^{\rm a}${}$^{\ast}$
\thanks{$^\ast$Corresponding author. Email: Claudio.Castelnovo@rhul.ac.uk
\vspace{6pt}} 
and Claudio Chamon$^{\rm b}$
\\\vspace{6pt} 
$^{\rm a}${\em{
SEPnet and Hubbard Theory Consortium, 
Department of Physics, 
Royal Holloway University of London,
Egham TW20 0EX, UK
}}\\ 
$^{\rm b}${\em{
Department of Physics, 
Boston University, 
Boston MA, 02215 USA
}}\\\vspace{6pt}\received{v1.0 released July 2011} }

\maketitle

\begin{abstract}
Quantum tunneling often allows pathways to relaxation past energy barriers 
which are otherwise hard to overcome classically at low temperatures. 
However, this is not always the case. 
In this paper we provide exactly solvable examples where the barriers 
each system encounters on its approach to lower and lower energy states 
become increasingly wide and eventually scale with the system size. 
If the environment couples {\it locally} to the physical degrees of
freedom in the system, tunnelling under these barriers requires processes 
whose order in perturbation theory is proportional to the width of the 
barrier. 
This results in quantum relaxation rates that are exponentially suppressed 
in system size: For these quantum systems, no {\it physical} bath 
can provide a mechanism for relaxation that is not dynamically 
arrested at low temperatures. 
The examples discussed here are drawn from three dimensional generalizations 
of Kitaev's toric code, originally devised in the context of topological 
quantum computing. They are devoid of any local order parameters or symmetry 
breaking and are examples of topological quantum glasses. 
We construct systems that have slow dynamics similar to either strong or 
fragile glasses. The example with fragile-like relaxation is interesting in 
that the topological defects are neither open strings or regular open 
membranes, but fractal objects with dimension $d^*=\ln 3/\ln 2$. 
\bigskip


\end{abstract}
%
%
%%%%%%%%%%%%%%%%%%%%%%%%%%%%%%%%%%%%%%%%%%%%%%%%%%%%%%%%%%%%%%%%%%%%%%%%%%%%%%%

\section{\label{sec: intro}
Introduction: How slow is slow?
        }
%brief history on glassiness and the difficulty of identifying a true 
%``glass transition'' : 
The problem of the approach to equilibrium is a formidable one. 
Nature provides us with plenty of examples of systems that simply do not 
equilibrate in any experimentally accessible times, starting from a 
material that is so commonly accessible to become synonym of an entire 
field of research: SiO$_2$, \textit{glass}~\cite{Angell1995}. 
A great deal of effort has been directed, both by physicists and chemists 
alike, to the question of how we can understand and describe these systems 
when they encounter dynamical obstructions in their attempted path to 
thermal equilibration. 
In spite of much research, however, a complete theoretical description of 
the glass transition remains an open problem. 

%why do we care about slow dynamics? 
Understanding the origin of long equilibration time scales is both of
fundamental significance from a theoretical point of view and of
great importance from a technological point of view. Uncovering the
reasons why systems fall out of equilibrium might enable us
to manipulate them, either hindering or enhancing the slow
dynamics depending on the specific application. Advancements on
understandying the properties of glassy materials would have
implications to problems ranging from biology to petroleum recovery.

Before we begin our discussion, one should decide on a working
definition of slow dynamics: \textit{how slow is slow?}  On the one
hand, one can set a threshold -- for instance, the famed viscocity of
$10^{13}$~Poise~\cite{Angell1995} -- beyond which a system is said to
be in a glassy state. A classification can then be developed based on
how quickly the time scales grow as a function of some natural tuning
parameter that leads the system into such state (e.g., temperature or
concentration). According to this classification, if the parameter is
temperature, one distinguishes between strong and fragile glasses if
the equilibration time scales grow in an Arrhenius form $\tau \sim
\exp[\Delta / T]$ or faster, respectively. 
Examples of the latter are the exponential inverse temperature square 
behavior, $\tau \sim \exp[\Delta^2 / T^2]$, and the Vogel-Fulcher law for a 
true glass transition at finite $T_0$, $\tau \sim \exp[\Delta /(T-T_0)]$.

On the other hand, one can classify systems \textit{in their glassy state} 
by looking at how fast their large time scales grow with system size. 
While in a finite system all time scales are strictly speaking 
finite,\footnote{Note however that these time scales may already be in 
practice longer than any experimentally accessible times.} 
they can exhibit a dependence on system size. 
In general, one can account for three different scenarios. 
(i) There could be no size dependence at all. For instance, 
a purely Arrhenius behaviour with a local energy barrier $\Delta$ can lead 
to time scales $\sim \exp(\Delta / T)$ that are independent of the size of 
the system. 
(ii) Time scales can grow polynomially in system size, 
$\tau \sim {\rm poly}(L)$. For example, this is the case of 
critical slowing down and diffusive modes. 
Finally, (iii) time scales can grow exponentially in system size, 
$\tau \sim \exp(L)$.\footnote{In any realistic 
(i.e., extensive) system, time scales cannot grow faster than exponential 
in the volume of the system.} 
[In the following we shall use the short hand notation $\tau \sim \exp(L)$ 
to denote generic exponential time scales of the form $\exp(a L^b)$.] 
Examples include the Sherrington-Kirkpatrick model~\cite{Sherrington1975} 
and $p$-spin glasses~\cite{Crisanti1992,Crisanti1993}. 

Drawing a line between time scales that grow polynomially vs. exponentially 
in system size can be understood within the framework of computational 
complexity: %~\cite{NielsenChuang_book}: 
Nature itself is a computer (which could in turn be simulated
in a universal digital computer), running its algorithm for dynamical
evolution. Any system thermalizes if the algorithm is allowed to run 
indefinitely. However, in finite times one has to come to terms with how 
efficient the dynamical evolution algorithm is. 
The system size $L$ is the size of the problem and 
systems whose equilibration times grow exponentially in $L$ correspond 
to computationally hard problems for nature's algorithm. 

%A characterisation of glasses based on system size dependence of their 
%equilibration time scales is also useful in order to address quantum 
%mechanical systems. 
%Determining how time scales grow as a function of some tuning parameter 
%$g$ in a quantum system is a tall order, and dynamical quantum simulations 
%are not readily available to the rescue of this analytical challenge. 
%On the contrary, investigating the properties of a quantum system in a 
%given phase is usually more affordable and one can study how long time 
%scales develop as the system size is varied. 

In this article we focus on the system size dependence of
equilibration time scales, in particular in quantum systems.  In
Sec.~\ref{sec: quantum glassiness}, we argue that any classical system
whose relaxation is exponential in $L$, when endowed with local
quantum dynamics, enters a quantum glass state (exponential in system
size). The converse is not true. Quantum glasses exist that `melt' at
any finite temperature, with the appearance of time scales that are
polynomial in system size. In Sec.~\ref{sec: quantum glass examples}
we provide three examples of this behaviour.

We note that exponentially long times naturally arise in systems undergoing 
spontaneous symmetry breaking. For example, going from 
positive to negative magnetization in the ordered phase of an Ising 
ferromagnet takes an exponentially large time in system size. 
These states however are distinguished by local order parameters (e.g., 
the local magnetization), which allow one to make significant progress in 
understanding their dynamics in terms of nucleation, domain growth and 
coarsening. 
In this paper we focus instead on systems where exponential relaxation 
time scales appear in the absence of local order parameters and symmetry 
breaking. 
At the quantum mechanical level, a natural context where to look for 
ground states without local order and without explicit disorder in the 
Hamiltonian or bath parameters is that of topologically ordered 
systems~\cite{Wen1990,Wen1995}. 
Indeed the examples of quantum glasses we put forward in 
Sec.~\ref{sec: quantum glass examples} are inspired by toy lattice models 
for topological order~\cite{Kitaev2003} and ought perhaps to be called 
\textit{topological quantum glasses}. 
A brief account on the physics of this family of quantum glasses, and 
in particular of the example in Sec.~\ref{sec:modelIb}, 
was discussed by one of the authors in Ref.~\onlinecite{Chamon2005}. 

Finally, in Sec.~\ref{sec: characterisation} we briefly comment on general 
approaches to characterise dynamical quantum systems and how -- due to the 
characteristic exponential dependence on system size -- one should beware 
of truncating perturbative approaches when studying quantum glasses, as 
they lead to reducible dynamics within disconnected sectors. 
%
%
%%%%%%%%%%%%%%%%%%%%%%%%%%%%%%%%%%%%%%%%%%%%%%%%%%%%%%%%%%%%%%%%%%%%%%%%%%%%%%%

\section{\label{sec: quantum glassiness}
Classical and quantum glasses
        }
It is often the case that quantum tunneling allows relaxation past energy 
barriers which are otherwise hard to overcome classically at low 
temperatures. 
For instance, the growth of equilibration time scales due to Arrhenius 
activated behaviour from a local, finite barrier $\tau \sim \exp(\Delta / T)$ 
is eventually cut off at sufficiently low temperatures by 
temperature-independent tunneling across the barrier, $\tau_q \sim \Delta/\t$ 
where $\t$ is some quantum mechanical amplitude for the 
process.\footnote{These time scales are dimensionless for convenience. 
They are intended as factors multiplying the characteristic microscopic 
time of the system in the absence of barriers. For instance, in the quantum 
mechanical tunnelling case, the characteristic time would be dictated by the 
inverse hopping amplitude, $1/t$.} 

In general, classical systems that enter a glass state exhibit equilibration 
time scales that grow exponentially in system size.\footnote{Systems where 
the glass transition occurs in the zero temperature limit ought to be 
treated with care, as the order of limits (thermodynamic vs. $T \to 0$) 
matters. For instance, time scales that are exponential in system size 
with a trivial Arrhenius activated prefactor $\exp(\Delta / T)$ diverge 
in the $T \to 0$ limit irrespective of system size.} 
The origin of such characteristic exponential dependence lies in the 
appearance of large energy barriers between the glassy free energy minima, 
whose height and width grow with the system size. 

What happens if we take a classical system with $\tau\sim\exp(L)$ and 
lower the temperature to zero, where coherent quantum mechanical processes 
become active? 
As we shall argue hereafter, the fact that the barriers grow with system 
size has a dramatic effect on the quantum mechanical relaxation time scales. 
Realistically, we assume that only local terms are allowed in 
the tunneling Hamiltonian -- for instance, a transverse field of magnitude 
$\t$ that flips individual spins in a localised spin system. 
Wide barriers mean that the system has to visit a large number 
$N_s \sim {\rm poly}(L)$ of excited states in its journey from one side to 
the other of the barrier. 
Quantum mechanically this process is strongly suppressed: 
$\tau_q \sim (U/\t)^{N_{s}}$, where $U$ is the energy
scale of the intermediate virtual states. 
Consequently, the quantum mechanical relaxation time scale acquires also a 
system size dependence of \textit{exponential form}, 
$\tau_q \sim \exp[N_s \ln(U/\t)] \sim \exp(L)$. 

We conclude that a classical model with $\tau\sim\exp(L)$ remains 
exponentially slow at $T=0$ even if (local) quantum tunneling processes are 
allowed. 
A whole family of quantum glasses can thus be derived directly from 
classical glassy systems by replacing temperature with local quantum 
dynamics. One might wonder whether the converse is true: 
Are there quantum glasses that do not have a classical parent? 
In other words, are there quantum systems that have exponential relaxation 
times $\tau_q \sim \exp(L)$ at $T=0$, but that immediately `melt' to 
$\tau \lesssim {\rm poly}(L)$ as soon as $T>0$? 
In Sec.~\ref{sec: quantum glass examples} we will answer positively to 
this question by explicit construction of local Hamiltonians that are 
examples of \textit{purely quantum} glasses. 
%
%
%%%%%%%%%%%%%%%%%%%%%%%%%%%%%%%%%%%%%%%%%%%%%%%%%%%%%%%%%%%%%%%%%%%%%%%%%%%%%%%

\section{\label{sec: quantum glass examples}
Examples of quantum glasses without disorder
        }
In order to construct our quantum Hamiltonians, we shall draw inspiration 
from two rather distinct areas: 
(i) classical kinetically constrained models, where slow dynamics appear 
without disorder; and 
(ii) spin models for topological order, which exhibit gapped quantum ground 
states devoid of any local order parameters or symmetry breaking. 

Kinetically constrained models allowed to make progress on the study of the 
classical glass transition by investigating systems in which the equilibrium 
properties are easy to understand, but the dynamics are non-trivial due to 
kinetic constraints (for a review, see Ref.~\onlinecite{Ritort2003}). 
Simple lattice models are constructed embodying the notion that there are 
jammed and unjammed regions in glass formers described by a 
discrete state, which is representable in terms of Ising spin 
variables~\cite{Garrahan2003}. 
One particularly interesting class is that of spin plaquette models with 
nontrivial classical energy function and unconstrained spin dynamics, 
which can be equivalently described in terms of non-interacting defect 
variables with rather constrained multi-defect 
dynamics~\cite{Newman1999,Garrahan2000,Buhot2002,Garrahan2002}. 
The thermodynamics of these systems is therefore trivial, while their 
dynamics is not, displaying a rich non-equilibrium behaviour 
(ageing, for instance). 

Topological order is a relatively recent concept in systems of strongly 
interacting particles~\cite{Wen1990,Wen1995}. 
Some quantum phases of matter, in contrast to common examples like crystals 
and magnets, are not characterized by a local order parameter and broken 
symmetries. Instead, 
they are characterized by a ground state degeneracy (when the system is 
defined on a torus or other surface of higher genus) that cannot be 
lifted by any local perturbation. This degeneracy is topological in nature 
and it is intimately related to quantum number fractionalization. 
The robustness of the topological degeneracy against local noise 
due to the environment is at the core of topological 
quantum computation~\cite{Kitaev2003}. %,NielsenChuang_book}. 

Strong correlations that lead to these exotic quantum spectral properties 
can also impose kinetic constraints similar to those studied in the context 
of glass formers. 
Here we presents concrete examples of systems that 
have local Hamiltonians with no quenched disorder, 
exactly solvable spectra and topologically ordered quantum ground states. 
In these examples, not only do the classical thermal barriers grow with 
decreasing temperature, but also the widths of the quantum 
tunneling barriers do, in such a way that quantum processes are even more 
severely suppressed than classical ones at low temperatures. 
The origin of this behavior is the fact that \textit{any} bath couples 
\textit{locally} to the physical degrees of freedom of the system and it 
can only flip large objects through virtual processes 
of large order in the system-bath coupling. For these quantum systems, no 
\textit{physical} bath can provide a mechanism for relaxation that is 
\emph{not} dynamically arrested at low temperatures.

We first discuss a two-dimensional (2D) 
quantum system with strong glass-like relaxation times when in contact with a 
restricted class of thermal baths; this example is used to clarify the 
issue of how a given Hamiltonian requires a minimum number of degrees of 
freedom that the bath must locally control for the system to be able to 
equilibrate. The second example is a three-dimensional quantum system with 
strong glass-like relaxation times for any class of baths that couple locally 
to physical degrees of freedom of the system. The third example is a 
three-dimensional quantum system with fragile glass-like relaxation times.
%
%
%-----------------------------------------------------------------------------

\subsection{\label{sec:modelI}
Warmup: 2D example
           }
\begin{figure}
\begin{center}
\resizebox*{8cm}{!}{\includegraphics{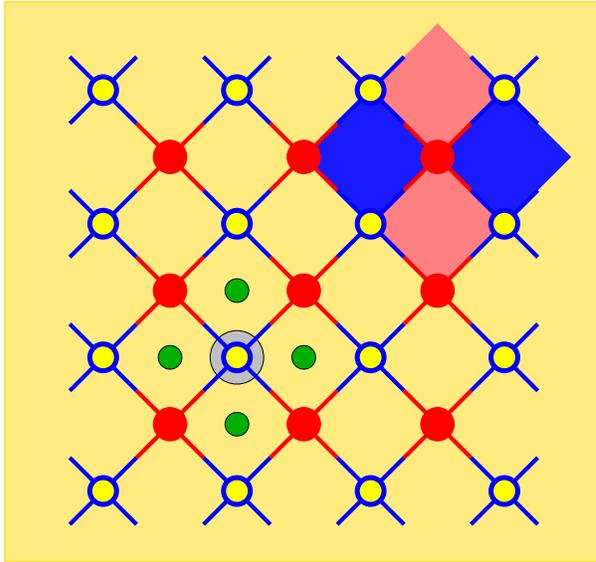}}
\caption{
Square lattice, with spin operators defined on the sites. The square lattice
is bipartite, and the two sets of points $A$,$B$ are shown in red solid dots 
and blue open circles. 
A diamond contains 4 vertices in an elementary plaquette, and the diamonds can
also be divided into two sets (forming a red/blue checkerboard) according to
which sublattice their topmost vertices belong to. Four-spin operators are
defined on each plaquette using the $\sigma^{\rm x}$ and $\sigma^{\rm y}$
components of the spin, as described in the text. The green dots
correspond to ``defects'' that are generated by applying a $\sigma^{\rm z}$
to the site encircled.
}
\label{fig:square} 
\end{center}
\end{figure}

The first model is constructed on a two-dimensional (2D) square lattice,
shown in Fig.~\ref{fig:square}. Each site can be labeled by $i,j\in \mathbb
Z$ that index a site in the Bravais lattice spanned by the primitive vectors
$\bm a_1=(1,1)/\sqrt{2}$ and $\bm a_2=(-1,1)/\sqrt{2}$. To shorten the 
notation, define a superindex $I\equiv (i,j)$. At every lattice site $I$ one
defines quantum spin $S=1/2$ operators $\sigma^{\rm x}_{I}$, 
$\sigma^{\rm y}_{I}$, and $\sigma^{\rm z}_{I}$. 
The square lattice is bipartite: it contains two sets of sites, which we 
label $A$ and $B$, and which are shown in red and blue color in 
Fig.~\ref{fig:square}.

Let us define the quantum Hamiltonian in terms of the spins 
$\bm \sigma_{I}$. Here we follow an approach similar to that of Kitaev, who
constructed in a beautiful paper model quantum Hamiltonians that are exactly
solvable~\cite{Kitaev2003}. In those models, the spins resided on links in
planar lattices, but it is possible to carry out similar constructions with
spins defined on vertices~\cite{Wen2003}, as it is done here. Later on we 
show how the construction with spins on vertices can be generalized to 3D
lattices.

Define a {\it diamond} cell $P_I$ as the set of four lattice sites in an
elementary plaquette with site $I$ at its top. The four vertices are indexed
by $J_n(I)$, for $n=1,\dots, 4$, with one of the vertices $J_1(I)=I$. The
four labels are assigned in such as way that the pairs
$\{J_1,J_3\},\{J_2,J_4\}$ are diagonally opposite sites from one another.
Explicitly,
$J_1(I)=I\equiv(i,j)$, $J_2(I)\equiv(i-1,j)$, $J_3(I)\equiv(i-1,j-1)$, and
$J_4(I)\equiv(i,j-1)$. It is simple to see that the total number of diamonds
equals the number of spins: each lattice site $I$ is the top vertex of a
single diamond. The one-to-one relation between a site $I$ and the diamond
$P_I$ allows us to partition diamonds into two sets $A$ and $B$ (and
color the corresponding diamonds red and blue, as shown in
Fig.~\ref{fig:square}).

Now define the operators ${\cal O}_{I}$ as
\begin{equation}
{\cal O}_{I}=
\sigma^{\rm y}_{J_1(I)}\;
\sigma^{\rm x}_{J_2(I)}\;
\sigma^{\rm y}_{J_3(I)}\;
\sigma^{\rm x}_{J_4(I)}
\;.
\label{eq:O-def}  
\end{equation}
These operators commute, $[{\cal O}_I,{\cal O}_{I'}]=0$ for all pairs $I,I'$.
It is simple to see how: two diamonds $P_I$ and $P_{I'}$ can share 0,1, or,
2 spins ($I \neq I'$). If they share 0 spins, they trivially commute. If they
share 1 spin, the component (x,y or z) of $\bm \sigma$ for that shared spin
coincides for both ${\cal O}_I$ and ${\cal O}_{I'}$ (the two diamonds touch
along one of their diagonals). If they share 2 spins, the components of 
$\bm \sigma$ used in the definition of ${\cal O}_I$ and ${\cal O}_{I'}$ are 
different for both spins, there is a minus sign from commuting the x and y
components of the spin operators from each of the shared spins, and the two
minus signs cancel each other.

Consider the system Hamiltonian
\begin{equation}
\hat H=-h\sum_I {\cal O}_I
\;,
\label{eq:Hamiltonian}  
\end{equation}
which is trivially written in terms of the ${\cal O}_I$ operators, but
complicated in terms of the original spins $\bm \sigma_I$. Because the 
${\cal O}_I$ commute, the eigenvalues of the Hamiltonian can be labeled 
by the list of eigenvalues $\{O_I\}$ of all the ${\cal O}_I$. 
Note that ${\cal O}_I^2=\openone$, and each $O_I=\pm 1$. 
In particular, the ground state corresponds to $O_I=1$ for all $I$.

Because the number of spins equals the number $N$ of sites, one may naively
expect that the list $\{O_I=\pm 1\}$ exhausts the $2^N$ states in the Hilbert
space, spanned by $\{\sigma^{\rm z}_I=\pm 1\}$. However, there are constraints
that the ${\cal O}_I$ satisfy when the system is subject to periodic boundary
conditions (compactified to a torus). One can show that
\begin{equation}
\prod_{I\in A} {\cal O}_I=
\prod_{I\in B} {\cal O}_I=
\openone
\;.
\label{eq:constraints}  
\end{equation}
There are two constraints; therefore there are only $2^{N-2}$ independent
$\{O_I=\pm 1\}$. This implies, in particular, that there is a ground state
degeneracy of $2^2=4$. (Notice that the ground state degeneracy is not
associated with a symmetry. In particular, it is easy to show that $\langle
\sigma^{\rm x,y,z}_I\rangle=0$.) This is a topological degeneracy and the
eigenvalues of a set of two non-local (winding or topological) operators
${\cal T}_{1,2}$ are needed to distinguish between the 4 degenerate ground
states.

The operators ${\cal T}_{1,2}$ can be constructed as follows. 
Let ${\cal P}_l=\{I \,|\, i+j=l\}$ be a set of points along a horizontal 
line. Notice that
sites on a line belong either all to sublattice $A$ or all to sublattice $B$,
for example ${\cal P}_{1}\subset A$ and ${\cal P}_{2}\subset B$. 
Define
%
%
%\begin{subequations}
%\begin{eqnarray}
%{\cal T}_{1}&=&\prod_{I\in {\cal P}_{1}}\sigma^{\rm y}_I\\
%{\cal T}_{2}&=&\prod_{I\in {\cal P}_{2}}\sigma^{\rm y}_I
%\;.
%\label{eq:calT}  
%\end{eqnarray}
%\end{subequations}
\begin{equation}
{\cal T}_{1}=\prod_{I\in {\cal P}_{1}}\sigma^{\rm y}_I
\qquad
{\cal T}_{2}=\prod_{I\in {\cal P}_{2}}\sigma^{\rm y}_I
.
\label{eq:calT}  
\end{equation}
It is simple to check that $[{\cal T}_{1,2},{\cal O}_I]=0$ for all $I$ and
the two operators ${\cal T}_{1,2}$ trivially commute. Hence the two
eigenvalues ${T}_{1,2}=\pm 1$ of ${\cal T}_{1,2}$ can distinguish the 4
degenerate ground states. This degeneracy has a topological origin and it 
scales with the genus of the surface. 

The spectrum of the Hamiltonian Eq.~(\ref{eq:Hamiltonian}) is that of a
trivial set of $N-2$ free spins, determined by the list of eigenvalues
$\{O_I=\pm 1\}$ of all the ${\cal O}_I$, subject to the condition
Eq.~(\ref{eq:constraints}): $E_{\{O_I\}}=-h\sum_I O_I$. Excitations
above the ground state ($O_I=1$ for all $I$) are ``defects'' where $O_I=-1$
in certain sites $I$. Because of the constraints in Eq.~(\ref{eq:constraints}),
the defects appear only in pairs. These defects have non-trivial quantum
statistics: they are Abelian anyons with statistical angle different from
that of fermions or bosons~\cite{Kitaev2003}.

The equilibrium partition function (within a topological sector) is given by
$Z=\sum_{\{O_I=\pm1\}} e^{\beta h\sum_I O_I}$. In thermal equilibrium at
temperature $T$, the thermal average 
$\langle O_I\rangle = \tanh \frac{h}{T}$, and the concentration or density 
of $O_I=-1$ defects is $c=\frac{1}{2}\left(1-\tanh \frac{h}{T}\right)$. 
Notice that we have encountered a situation analogous to 
classical spin facilitated models~\cite{Ritort2003}, 
in particular the plaquette models displaying glassy 
dynamics~\cite{Newman&Moore99,Garrahan&Newman00,Buhot&Garrahan02,Garrahan2002}:
the thermodynamics is trivial in terms of non-interacting defect variables. 

What about the dynamics of our quantum model?
Although the spectrum of the model we are discussing is the same as that of
free spins in a uniform magnetic field $h$, the variables $O_I$ for different
diamonds $I$ {\it cannot} be independently changed. 
In fact, the operators ${\cal O}_I$ involve four spins, shared by neighboring 
operators ${\cal O}_{I'}$. Under the action of any {\it local} spin operator, 
one cannot change the eigenvalue of $O_I$ without changing the eigenvalues 
$O_{I'}$ of its neighbors. 
What are the allowed, {\it physical} dynamical evolution rules for 
this quantum system? Can these dynamical rules lead to equilibration?

In order to endow the system with some physical dynamics, we couple
the original physical spins to {\it individual} baths at temperature
$T$. Here we do not consider ``Turkish'' baths of multiple spins; 
nonetheless, as long as the groups of spins sharing a bath are locally
delimited in space, the results obtained below should remain qualitatively
unchanged. Moreover, allowing the bath to communicate information
through long-ranged couplings (via phonons, for instance) will not
change the results, as long as it operates on delimited regions of
space. %at both ends that it acts on. 

When the original individual spins are coupled to their baths, ``flips'' of
the states of multiple ${\cal O}_I$ sharing a given spin take
place. Therefore, our model has a trivial spectrum but a highly correlated
dynamics. It is this correlated dynamics that gives rise to non-trivial
non-equilibrium behaviour.

We introduce the bath degrees of freedom as in the
Feynman-Vernon influence functional approach~\cite{Feynman-Vernon63} or
Caldeira-Leggett dissipative quantum mechanics
formulation~\cite{Caldeira-Leggett81,Caldeira-Leggett83}, by letting the
Hamiltonian of the system plus bath be
$\hat {\cal H} = \hat H+ \hat H_{\rm bath}+\hat H_{\rm spin+bath}$, where
$\hat H$ is defined in Eq.~(\ref{eq:Hamiltonian}), and
\begin{subequations}
\begin{eqnarray}
&&\!\!\!\!\!\!\!\!\!\hat H_{\rm bath}=\sum_{I,\alpha} \int_0^\infty \!\!\! dx
\;[\Pi^\alpha_I(t,x)]^2 + [\partial_x \Phi^\alpha_I(t,x)]^2 
\label{eq:Hamiltonian-bath}
\\
&&\!\!\!\!\!\!\!\!\!\hat H_{\rm spin/bath}= 
\sum_{I,\alpha} g_\alpha\; \sigma^\alpha_I \; \Pi^\alpha_I(t,0)
\,.
\label{eq:Hamiltonian-spin+bath}
\end{eqnarray}
\end{subequations}
The three components ($\alpha=1,2,3$) of the conjugate vector
fields $\bm\Phi_I$ and $\bm\Pi_I$ obey the equal-time commutation relation
$
[\Phi^\alpha_I(t,x),\Pi^{\alpha'}_J(t,x')]=i\;\delta_{IJ}\;
\delta_{\alpha\alpha'} \; \delta(x-x')
$. 

Notice that, for each site $I$, the bath-spin system can be viewed as an
extended bosonic string that couples to a spin at the boundary $x=0$. The
coupling amplitudes are $g_\alpha$. One can in general choose anisotropic
couplings, but the most general bath should contain all of $g_{1,2,3}$. In
the quantum model, acting on a site $I'\in P_I$ with one of 
$\sigma_{I'}^{\rm x},\sigma_{I'}^{\rm y}$, or $\sigma_{I'}^{\rm z}$ flips 
or not the eigenvalue $O_I$ depending on whether 
$\sigma_{I'}^{\rm x,y,z}\;{\cal O}_I=\mp{\cal O}_I\;\sigma_{I'}^{\rm x,y,z}$, 
respectively.

If integrated out, the bath degrees of freedom away from the boundary $x=0$
lead to a non-local in time action and to dissipation effects. Instead of
working with the dissipative action, let us follow the time evolution of the
system plus bath and look at the possible evolution pathways of the quantum
mechanical amplitudes of the system plus bath degrees of freedom. After
evolution by time $t$ from some initial state, the system is in a quantum
mechanical superposition
\begin{equation}
\vert\Psi\rangle 
=
\sum_{\{O_I=\pm 1\}} \Gamma_{\{O_I\}}\;\vert{\{O_I\}}\rangle
\otimes\vert\Upsilon_{\{O_I\}}\rangle
\,,
\label{eq:state}
\end{equation}
where $\vert\Upsilon_{\{O_I\}}\rangle$ is a state in the bath Hilbert space 
with norm one. 
(Here we focus on states in a single topological sector, although
mixing sectors can be done by including the eigenvalues of the topological
operators ${\cal T}$; mixing is exponentially suppressed as the system size
increases). The fact that the bath degrees of freedom couple to single
quantum spins $\bm \sigma_I$ enters in the problem through the permitted
channels for transferring amplitudes among the $\Gamma_{\{O_I\}}$.

The processes that transfer amplitude among the $\Gamma_{\{O_I\}}$ correspond
to different orders in perturbation theory on the system-bath coupling 
$g_\alpha$. There is also a thermal probability factor coming from the bath 
and that depends on the difference between the initial and final energy
$E_{\{O_I\}}=-h\sum_I O_I$ of the system. One class of paths is a {\it
sequential} passage over states connected through order $g_\alpha$ processes;
this is a ``semi-classical'' type trajectory. 

Within this restricted class of processes, we can make a connection to the
classical plaquette models, particularly the $4$-spin square plaquette model
whose glassy properties have been studied in 
Refs.~\onlinecite{Alvarez-Franz-Ritort96,Lipowski97,Buhot&Garrahan02}. 
This classical model is obtained by defining diamond variables 
$\tau_I$ in place of the ${\cal O}_I$ as in Eq.~(\ref{eq:O-def}) using 
{\it only}, say, the z-component $\sigma^{\rm z}$ for {\it all} four sites 
of the diamonds. 
Flipping an individual spin changes signs to all four $\tau_I$ variables 
surrounding the spin. This multi-defect type dynamics makes it
difficult for the system to relax to equilibrium. For example, if the
temperature is lowered, in order to decrease the defect density, either four
defects come together and annihilate ($4\to 0$ decay), or three defects
become one ($3\to 1$ decay). However, the defects are not free to 
diffuse and come together. In order to move an isolated defect, it must first
decay into three defects ($1\to 3$ production), then a pair can diffuse 
freely, and recombine with another defect elsewhere through a $3\to 1$ decay 
process. The final result is that the original defect (as well as the other 
defect elsewhere) moves by one lattice spacing and the total number of 
defects does not change (it first increases by $2$ and then decreases by $2$). 
Because of the initial $1\to 3$ production process, there is an energy barrier 
of $2h$ to overcome. This activation barrier leads to 
recombination/equilibration times 
\begin{equation}
t_{\rm seq.}\sim\exp(2h/T) 
\label{eq: tau c}
\end{equation}
that grow as temperature is lowered in an Arrhenius 
fashion~\cite{Buhot&Garrahan02}.

In our model, the same situation is recovered if the coupling to the bath 
involves {\it only} $\sigma^{\rm z}$ components of the spins (i.e., 
$g_1=g_2=0$). The Hamiltonian encompasses only $\sigma^{\rm x}$ and 
$\sigma^{\rm y}$ components and the action of a $\sigma^{\rm z}$ operator 
on a spin changes the eigenvalues of all four surrounding ${\cal O}_I$. 
One may then wonder how quantum tunnelling processes modify the 
behaviour of the system as compared to its classical counterpart. 
Recall that defect annihilation occurs only through virtual
processes in which the number of defects is strictly larger in the 
intermediate (virtual) steps. At temperature $T$, the typical defect 
separation is $\xi=c^{-1/2}\sim e^{h/2T}$. Bringing them together requires 
tunnelling processes at least of order $\xi$ in perturbation theory, which 
have an amplitude of order $(g/h)^\xi$ (notice the energy denominator $h$). 
This amplitude leads to recombination/equilibration times 
\begin{equation}
t_{tun.} 
\sim 
\exp\left[\ln(h/g)\;e^{h/2T}\right] 
\,,
\label{eq: tau q}
\end{equation}
which grow extremely fast as the temperature is lowered. 
What we learn from this simple estimation is that
quantum tunneling is less effective than classical sequential processes in
thermalizing the system. This is counterintuitive to the notion that at low
temperatures quantum tunneling under energy barriers remains an open process
while classical mechanisms are suppressed due to thermal activation
costs. The reason for this particular quantum freezing is simple: 
equilibration proceeds through activated defect recombination; as the
density of defects decreases at low temperatures, the barrier {\it widths}
increase and debilitate quantum tunneling. 
In passing, we note that in a finite system of size $L$, one must replace 
$\xi$ by $L$ in the estimation of the recombination/equilibration times, 
$t_{tun.}\sim \exp\left[\ln(h/g)\;L\right]$; this time scale is also of the 
order of that for tunneling between two topological ground states in a 
finite system of size $L$~\cite{Kitaev2003}. 

Let us return to the issue of which component of spin enters in the coupling
to the bath. The minimal bath coupling involves the z-component of the spin 
operator, for this component does not commute with {\it any} of the 
${\cal O}_I$ that contain that given spin. 
However, there is no \textit{a priori} reason why the coupling should be 
restricted to z-components only. 
If the bath couples to x- and y-components as well, defects can diffuse 
freely without barriers (via $1 \to 1$ processes), eliminating the need for 
$1\to 3$ defect production processes in order to bring them together and 
annihilate them. 
Simple defect diffusion leads to a diffusive equilibration time 
$t_{eq}\sim L^2$ for a system of size $L$ (polynomial in $L$ with 
constant exponent). 

Such dependence on the details of the bath coupling is removed in the 
3D models we discuss next, one of which has even slower equilibration, 
as in fragile glasses. 
%
%
%------------------------------------------------------------------------------

\subsection{\label{sec:modelIb}
3D quantum strong glass
           }
In the previous example, particle diffusion finds its origin in the fact that 
the x- or z-components of a spin commute with 2 out of the 4 surrounding 
diamonds. Therefore, defects can be created in pairs, not quadruplets, and 
single defect diffusion can take place through annihilation of a defect and 
creation of a neighbouring one via pair flip. In the example that 
follows (see also Refs.~\cite{Chamon2005,Bravyi2011}), 
six defect cells share each single spin, in such a way that acting 
with any component of the spin operators flips 4 cells and defect 
diffusion cannot occur.

The model displaying strong like glassiness is constructed on a
three-dimensional (3D) face-centered cubic (fcc) Bravais lattice, spanned by
the primitive vectors $\bm a_1=(1,1,0)/\sqrt{2}$, $\bm a_2=(0,1,1)/\sqrt{2}$, 
and $\bm a_3=(1,0,1)/\sqrt{2}$.  
%$\bm a_1=\frac{1}{\sqrt{2}}(1,1,0)$, $\bm a_2=\frac{1}{\sqrt{2}}(1,0,1)$, and
%$\bm a_3=\frac{1}{\sqrt{2}}(0,1,1)$.
Each site can be indexed by $i,j,k\in \mathbb Z$, and to shorten the
notation, define a superindex $I\equiv (i,j,k)$. At every lattice site $I$
one defines quantum spin $S=1/2$ operators $\sigma^{\rm x}_{I}$, 
$\sigma^{\rm y}_{I}$, and $\sigma^{\rm z}_{I}$.

\begin{figure}
\begin{center}
\resizebox*{8cm}{!}{\includegraphics{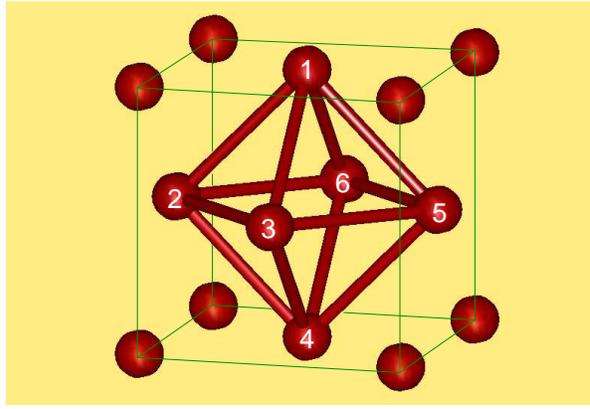}}
\caption{
Cubic cell of an fcc lattice. The centers of the six faces form an
octahedron, with its sites labeled from 1 (topmost) to 6. In addition to the
set of octahedra formed by the face centered sites, there are three more sets
of octahedra that can be assembled from sites both on faces and on corners of
the cubic cells, totalling 4 such sets. Six-spin operators are defined on
these octahedra using the $\sigma^{\rm x,y,z}$ components of spin on each
vertex as described in the text.
}
\label{fig:fcc} 
\end{center}
\end{figure}

The fcc lattice hosts sets of {\it octahedra}: the simplest one to
visualize is the one assembled from the centers of the six faces of a cubic
cell, as shown in Fig.~\ref{fig:fcc}. In addition to this simple
set, there are three more sets of octahedra that can be assembled from sites
both on faces and on corners of the cubic cells, totalling 4 such sets, which
we label $A,B,C$ and $D$.

It is simple to see that the total number of octahedra equals the number of
spins: each lattice site $I$ is the topmost vertex of a single octahedron.
Define then $P_I$ as the set of six lattice points in the
octahedron with site $I$ at its top. The six vertices are indexed by
$J_n(I)$, for $n=1,\dots, 6$, with one of the vertices $J_1(I)=I$. The six
labels are assigned in such as way that the pairs $\{J_1,J_4\}$, 
$\{J_2,J_5\}$, $\{J_3,J_6\}$ are diagonally opposite sites from one 
another. This number labeling is illustrated for a single octahedron in
Fig.~\ref{fig:fcc}. From the one-to-one relation between a site $I$ and the
octahedra $P_I$, we can also partition the lattice sites into the four sets
$A,B,C$ and $D$ of octahedra.

Now define the operators ${\cal O}_{I}$ as
\begin{equation}
{\cal O}_{I}=
\sigma^{\rm z}_{J_1(I)}\;
\sigma^{\rm x}_{J_2(I)}\;
\sigma^{\rm y}_{J_3(I)}\;
\sigma^{\rm z}_{J_4(I)}\;
\sigma^{\rm x}_{J_5(I)}\;
\sigma^{\rm y}_{J_6(I)}
\;.
\label{eq:O-octa-def}  
\end{equation}
These operators commute $[{\cal O}_I,{\cal O}_{I'}]=0$ for all pairs $I,I'$.
Indeed, two distinct octahedra $P_I$ and $P_{I'}$ can either 
share 0,1, or, 2 spins. If they share 0 spins, they trivially commute. If
they share 1 spin, the component (x,y or z) of $\bm \sigma$ for that shared
spin coincides for both ${\cal O}_I$ and ${\cal O}_{I'}$ (the two octahedra
touch along one of their diagonals). If they share 2 spins, the components
$\bm \sigma$ used in the definition of ${\cal O}_I$ and ${\cal O}_{I'}$ are
different for both spins. Thus, there is a minus sign from commuting the spin 
operators from each of the shared spins, and the two minus signs cancel. 

Consider the system Hamiltonian as in Eq.~(\ref{eq:Hamiltonian}), which is
trivially written in terms of the ${\cal O}_I$ operators, but complicated in
terms of the original spins $\bm \sigma_I$. Because the ${\cal O}_I$ 
commute, the eigenvalues of the Hamiltonian can be labeled by the list of
eigenvalues $\{O_I\}$ of all the ${\cal O}_I$. Notice that ${\cal
  O}_I^2=\openone$ and so each $O_I=\pm 1$. In particular, the ground state
corresponds to $O_I=1$ for all $I$.

Because the number of spins equals the number $N$ of sites, one may naively
expect that the list $\{O_I=\pm 1\}$ exhausts the $2^N$ states in the Hilbert
space. However, there are constraints that the ${\cal O}_I$ satisfy when the
system is subject to periodic boundary conditions (compactified). One can
show that
\begin{equation}
\prod_{I\in A} {\cal O}_I=
\prod_{I\in B} {\cal O}_I=
\prod_{I\in C} {\cal O}_I=
\prod_{I\in D} {\cal O}_I=
\openone
\;.
\label{eq:constraints-3D-fcc}  
\end{equation}
There are four constraints; therefore there are only $2^{N-4}$ independent
$\{O_I=\pm 1\}$ and there is a ground state degeneracy of $2^4=16$ (for a 
thorough discussion of this degeneracy and its dependence on boundary 
conditions, see Ref.~\onlinecite{Bravyi2011}). 
This is a topological degeneracy and the number of ground states scales with 
the genus of the manifold the system is defined on. 
The eigenvalues of a set of four non-local (topological) operators 
${\cal T}_{1,2,3,4}$ are needed to distinguish between the 16 degenerate 
ground states. 

The operators ${\cal T}_{1,2,3,4}$ can be constructed as follows. Let 
${\cal P}_l=\{I \,|\, j+k=l\}$ be a set of points along a horizontal plane. 
Notice that
each plane contains sites in only two of the four sublattices $A,B,C,D$. For
example ${\cal P}_{1}\subset A\cup B$ and ${\cal P}_{2}\subset C\cup D$.
Define
%
%
%\begin{subequations}
%\begin{eqnarray}
%{\cal T}_{1}&=&\prod_{I\in {\cal P}_{1}\cap A}\sigma^{\rm z}_I\\
%{\cal T}_{2}&=&\prod_{I\in {\cal P}_{1}\cap B}\sigma^{\rm z}_I\\
%{\cal T}_{3}&=&\prod_{I\in {\cal P}_{2}\cap C}\sigma^{\rm z}_I\\
%{\cal T}_{4}&=&\prod_{I\in {\cal P}_{2}\cap D}\sigma^{\rm z}_I
%\;.
%\label{eq:calToctahedra}  
%\end{eqnarray}
%\end{subequations}
\begin{equation}
{\cal T}_{1}=\prod_{I\in {\cal P}_{1}\cap A}\sigma^{\rm z}_I
\qquad
{\cal T}_{2}=\prod_{I\in {\cal P}_{1}\cap B}\sigma^{\rm z}_I
\qquad
{\cal T}_{3}=\prod_{I\in {\cal P}_{2}\cap C}\sigma^{\rm z}_I
\qquad
{\cal T}_{4}=\prod_{I\in {\cal P}_{2}\cap D}\sigma^{\rm z}_I
\;.
\label{eq:calToctahedra}  
\end{equation}
It is simple to check that $[{\cal T}_{1,2,3,4},{\cal O}_I]=0$ for all $I$,
and the ${\cal T}_{1,2,3,4}$ trivially commute among themselves. Hence the
four eigenvalues ${T}_{1,2,3,4}=\pm 1$ of ${\cal T}_{1,2,3,4}$ can
distinguish the 16 degenerate ground states. 

In this model one can verify that, whichever component of spin enters
in the coupling to the bath, it is impossible to flip only a pair of defects
and thus there is no mechanism for defect diffusion. The reason is that any
site is shared by 6 octahedra, and the operators ${\cal O}_I$ for these cells
are such that one can divide the 6 into 3 groups of 2 octahedra that will
have in their definitions, respectively, the x, y, and z component of spin
operators at the shared site. Acting with either of the three components of
the spin operator on this shared site will flip at least four defects. 

Quantum glassiness, i.e., the behaviour in Eq.~(\ref{eq: tau q}) 
leading to $\tau_q \sim \exp(L)$, is thus protected against any local bath. 
On the other hand, as soon as we leave the $T=0$ limit, the system size 
dependence of the relaxation time scales changes radically. 
Indeed, the energy barrier to defect diffusion is only a `one-step' process. 
As soon as $1 \to 3$ decays are allowed to take place thermally, any two of 
the three new defects can diffuse freely across the system. The time scales 
for diffusion are simply rescaled by the factor $t_{\rm seq.}\sim\exp(2h/T)$, 
Eq.~(\ref{eq: tau c}), but the system size dependence remain 
only $\tau_c \sim {\rm poly}(L)$. 
%
%
%------------------------------------------------------------------------------

\subsection{\label{sec:modelII}
3D quantum fragile glass
           }
\begin{figure}
\begin{center}
\resizebox*{8cm}{!}{\includegraphics{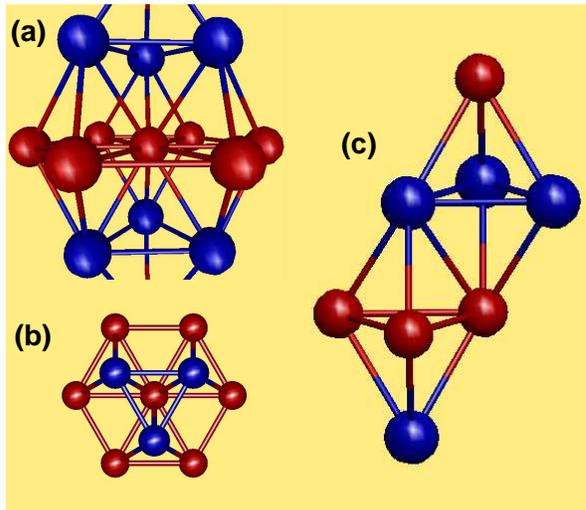}}
\caption{
Sites of an hcp (hexagonal close-packed) lattice. 
(a) The hcp lattice is comprised of two interpenetrating hexagonal lattices, 
which can be alternatively seen as stacked triangular lattices, shown in 
red and blue. Prisms are defined as sets of five sites, two of which belong 
to one sublattice (top and bottom of the prism), and three of which belong to 
the other and form a triangle that lies in the layer in between the top and 
bottom sites of the prism. Five-spin interactions are defined on each prism 
as explained in the text. 
(b) Top view of the hcp lattice, which shows that the blue sites stack on 
top of the red upward pointing triangles, and the red sites stack on top of 
the downward pointing blue triangles. 
(c) Two prisms with topmost sites belonging to different sublattices can 
share a common edge and the five-spin operators defined on the two prisms 
commute because minus signs from commuting the $\sigma^{\rm x}$ and 
$\sigma^{\rm z}$ components appear twice, once for each shared site, 
and cancel. 
}
\label{fig:prism}
\end{center}
\end{figure}

The model displaying fragile-like glassiness is constructed on a
three-dimensional (3D) hexagonal close-packed lattice, shown in
Fig.~\ref{fig:prism}. 
The lattice can be viewed as two interpenetrating
simple hexagonal Bravais lattices displaced from one another by
$\bm a_1/3 + \bm a_2/3 + \bm a_3/2$, where $\bm a_1=(1,0,0)$, 
$\bm a_2=(1,\sqrt{3},0)/2$, and $\bm a_3=(0,0,1)$ are the primitive vectors 
of the simple hexagonal lattice. 
The sites belonging to the two intercalating lattices are
shown in red and blue color in Fig.~\ref{fig:prism}. Each site can be
labeled by $i,j,k\in \mathbb Z$ that index a site in the Bravais lattice
spanned by $\bm a_{1,2,3}$, plus $q=0,1$ that indexes each of the two
sublattices -- to shorten the notation, define a superindex $I\equiv
(i,j,k;q)$. At every lattice site $I$ one defines quantum spin $S=1/2$
operators $\sigma^{\rm x}_{I}$, $\sigma^{\rm y}_{I}$, and 
$\sigma^{\rm z}_{I}$.

Define now a {\it prism} cell $P_I$ that contains five lattice sites
$J_n(I)$, for $n=1,\dots, 5$ as follows.  For a given lattice site $I$, the
prism $P_I$ contains the site $J_1(I)=I$, which belongs to one sublattice of
the hexagonal close-packed lattice, the three sites that belong to the other
sublattice and that form an elementary triangle (sites $J_2,J_3,J_4$) just
below the site $I$, and finally the site $J_5(I)$ just below that triangle,
which belongs to the same sublattice as site $I$. [In terms of the lattice
indices, $J_5(I)\equiv (i,j,k-1;q)$ if $I\equiv (i,j,k;q)$.]  An example of
two prisms is shown in Fig.~\ref{fig:prism}c. Notice that the two prisms
shown share a common edge, and that their tops belong to distinct (red and
blue) sublattices. The total number of prisms equals
the number of spins: each lattice site $I$ is the top vertex of a single
prism.

Now define the operators ${\cal O}_{I}$ as
\begin{equation}
{\cal O}_{I}=
\sigma^{\rm z}_{J_1(I)}\;
\sigma^{\rm x}_{J_2(I)}\;
\sigma^{\rm x}_{J_3(I)}\;
\sigma^{\rm x}_{J_4(I)}\;
\sigma^{\rm z}_{J_5(I)}
\;.
\label{eq:O-prism-def}  
\end{equation}
The operators commute, $[{\cal O}_I,{\cal O}_{I'}]=0$, for all pairs $I,I'$. 
Indeed, if $I,I'$ belong to the same sublattice and
the prisms $P_I,P_{I'}$ share a vertex, then they trivially commute as they
both involve the same component (x or z) of the spin operators $\bm \sigma$
at the shared site.  If they belong to distinct sublattices, they either
share 0 spins or an edge with 2 spins, as shown in Fig.~\ref{fig:prism}. If
they share 2 spins, the minus signs from commuting the x and z components of
spin in each of the shared sites appear an even number of times.

Consider the system Hamiltonian as in Eq.~(\ref{eq:Hamiltonian}), which is
trivially written in terms of the ${\cal O}_I$ operators, but complicated in
terms of the original spins $\bm \sigma_I$. Because the ${\cal O}_I$ 
commute, the eigenvalues of the Hamiltonian can be labeled by the list of
eigenvalues $\{O_I\}$ of all the ${\cal O}_I$. Notice that 
${\cal O}_I^2=\openone$, and so each $O_I=\pm 1$. 
In particular, the ground state corresponds to $O_I=1$ for all $I$.

Because the number of spins equals the number $N$ of sites, one may naively
expect that the list $\{O_I=\pm 1\}$ exhausts the $2^N$ states in the Hilbert
space. However, there are constraints that the ${\cal O}_I$ satisfy when the
system is subject to periodic boundary conditions (compactified). Each of the
two sublattices ($q=0,1$) of the hexagonal close-packed structure can be
further subdivided into $A_q,B_q$ or $C_q$ according to the three sublattices
of the tripartite triangular stacks of the simple hexagonal lattice. 
(All in all, there are six sublattices $A_{0,1},B_{0,1}$ and $C_{0,1}$.) 
One can show that
\begin{equation}
\prod_{I\in A_q\cup B_q} {\cal O}_I=
\prod_{I\in B_q\cup C_q} {\cal O}_I=
\prod_{I\in C_q\cup A_q} {\cal O}_I=
\openone
\;.
\label{eq:constraints-3D}  
\end{equation}
These are six constraints, but only four are independent, because the
product of the three products in Eq.~(\ref{eq:constraints-3D}) for the same
$q$ is trivially the identity. Therefore there are only $2^{N-4}$ independent
$\{O_I=\pm 1\}$. This implies, in particular, that there is a ground state
degeneracy of $2^4=16$ which is topological in nature and the number of 
ground states scales with the genus of the manifold the system is defined on. 
The eigenvalues of a set of four 
non-local (topological) operators ${\cal T}_{1,2,3,4}$ are needed to 
distinguish between the 16 degenerate ground states.

The operators ${\cal T}_{1,2,3,4}$ can be constructed as follows. Let the
plane ${\cal P}_{k,q}$ be the set containing sites with fixed $k$ and
$q$. Let
%
%
%\begin{subequations}
%\begin{eqnarray}
%{\cal T}_{1}&=&\prod_{I\in {\cal P}_{1,0}\cap (A_0\cup B_0)}\sigma^{\rm z}_I\\
%{\cal T}_{2}&=&\prod_{I\in {\cal P}_{1,0}\cap (B_0\cup C_0)}\sigma^{\rm z}_I\\
%{\cal T}_{3}&=&\prod_{I\in {\cal P}_{1,1}\cap (A_1\cup B_1)}\sigma^{\rm z}_I\\
%{\cal T}_{4}&=&\prod_{I\in {\cal P}_{1,1}\cap (B_1\cup C_1)}\sigma^{\rm z}_I
%\;.
%\label{eq:calTprisms}  
%\end{eqnarray}
%\end{subequations}
\begin{subequations}
\begin{eqnarray}
{\cal T}_{1}=\prod_{I\in {\cal P}_{1,0}\cap (A_0\cup B_0)}\sigma^{\rm z}_I
\qquad
{\cal T}_{2}=\prod_{I\in {\cal P}_{1,0}\cap (B_0\cup C_0)}\sigma^{\rm z}_I
\\
{\cal T}_{3}=\prod_{I\in {\cal P}_{1,1}\cap (A_1\cup B_1)}\sigma^{\rm z}_I
\qquad
{\cal T}_{4}=\prod_{I\in {\cal P}_{1,1}\cap (B_1\cup C_1)}\sigma^{\rm z}_I
\;.
\end{eqnarray}
\label{eq:calTprisms}  
\end{subequations}
It is simple to check that $[{\cal T}_{1,2,3,4},{\cal O}_I]=0$ for all $I$,
and the ${\cal T}_{1,2,3,4}$ trivially commute among themselves. Hence
the four eigenvalues ${T}_{1,2,3,4}=\pm 1$ of ${\cal T}_{1,2,3,4}$ can
distinguish the 16 degenerate ground states.

There are relations between this 3D model and a 2D classical triangular
plaquette model which has glassy
behavior~\cite{Newman&Moore99,Garrahan&Newman00,Garrahan2002}. The 2D
triangular plaquette model has Ising spin variables defined on the sites of a
triangular lattice and a 3-spin interaction which is the product of the
Ising variables on the downward pointing triangular plaquettes only. 
Plaquette Ising variables (the 3-spin products) are defined at the centre 
of the downward triangles, which behave thermodynamically as free Ising spins 
in a magnetic field. However, the dynamics is rather non-trivial in terms 
of the plaquette variables, for flipping an original spin corresponds 
to flipping all three surrounding plaquettes.

In our 3D model, each quantum spin $\bm \sigma_I$ is shared by 5 prisms: 3
whose centers are on the same plane, and 2 whose centers are immediately
above and below site $I$. If the system-nath coupling contains the
$\sigma^{\rm y}$ spin component, all 5 prisms are flipped. 
The $\sigma^{\rm z}$ and $\sigma^{\rm x}$ components flip either the 
eigenvalues of the 3 prisms on the plane or the 2 prisms on the vertical 
direction, respectively. 
Flipping the eigenvalues of 2 prisms in the vertical direction leads to
defect diffusion, but only in that direction. 

To connect our 3D quantum model to the triangular plaquette model, consider a
compactified slab (periodic boundary conditions) in the third dimension
(parallel to $\bm a_3$), with $M$ layers. Because of the periodic boundary
conditions, the odd-even parity of the defect number is conserved along
vertically stacked prisms regardless of the system-bath spin-flip operator,
$\sigma^{\rm x}$, $\sigma^{\rm y}$, or $\sigma^{\rm z}$. The defect number
parity can be captured by defining the following operator 
(recall $I \equiv(i,j,k;q)$): 
\begin{equation}
\tau_{i,j;q}=\prod_{k}{\cal O}_{(i,j,k;q)}
\label{eq:tau}
\;.
\end{equation}
It is also useful to define a similar product over the third dimension for
the original spins:
\begin{equation}
s_{i,j;q}=
\prod_{k} \sigma^{\rm x}_{(i,j,k;q)} 
\;.
\label{eq:s-def}  
\end{equation}
These ``slab'' operators allows us to concentrate on subspaces of the Hilbert
space with a given set of ${\tau_{i,j;q}}$ instead of the states with given
${O_{i,j,k;q}}$. There are dynamical processes that transfer quantum 
mechanical amplitudes within and between these subspaces labeled by 
${\tau_{i,j;q}}$; we can argue that the system is glassy by simply looking 
at the processes that transfer amplitude between the subspaces.

The variables ${\tau_{i,j;q}}$ and ${s_{i,j;q}}$ can effectively be used to
relate our quantum model to two 2D systems ($q=0$ or red, and $q=1$ or blue)
defined on sites labeled by $(i,j;q)$ of two distinct triangular lattices.
The variables $s_{i,j;q}$ can be related to the original spin variables in
the models of Refs.~\cite{Newman&Moore99,Garrahan&Newman00,Garrahan2002}. In
particular, one can relate the $s_{i,j;q}$ and the $\tau_{i,j;q}$ using
Newton's binomial coefficients through
\begin{equation}
s_{i,j;q}=\prod_{mn} [\tau_{n,m;q}]^{\binom{j-n}{i-m}}
\;.
\label{eq:s-tau}
\end{equation}
Using the Pascal triangle relation
${\binom{j+1-n}{i+1-m}}={\binom{j-n}{i-m}}+{\binom{j-n}{i+1-m}}$
one can show that indeed the defect variables correspond to
\begin{equation}
\tau_{i,j;q}=s_{i,j;q}\;s_{i+1,j;q}\;s_{i,j+1;q}
\;.
\label{eq:tau-s}
\end{equation}

We can focus again on two classes of processes: sequential passage over
states connected through order $g_\alpha$ processes (``semi-classical'' type
trajectories), and quantum tunneling process. The analysis of the
sequential processes is similar to that of the classical 2D
models~\cite{Newman&Moore99,Garrahan&Newman00,Garrahan2002} and goes as
follows. Single flips of $s$ correspond to concomitant flips of three $\tau$
defects. Defects can only be annihilated in triplets. The defects are not
free to diffuse and come together; instead, they move through the production
of more defects. For example, a defect can decay into two more defects, by
flipping one $s$ variable. Now, in order to bring three defects separated by
a distance $\xi$ together, one has to go through intermediate steps with a
large number of defects that are created. There is a hierarchical 
organization of these intermediate processes; equilateral triangles of size
$\xi=2^\ell$ require the creation of $\ell$ extra intermediate defects. Hence
there is an energy barrier of order $\ell h$ to be overcome. For a typical
equilibrium separation $\xi=c^{-1/2}\sim e^{h/2T}$, the barriers to be
overcome in the equilibration processes are of order $h^2 / (T \, 2\ln 2)$. 
Hence, the equilibration time scales as 
\begin{equation}
t_{\rm seq.}\sim\exp\left[\frac{(h/T)^2}{2\ln 2}\right]
\,,
\label{eq: tau c fragile}
\end{equation}
a much slower relaxation than the Arrhenius one, Eq.~(\ref{eq: tau c}), 
for the model discussed in Section~\ref{sec:modelIb}.

\begin{figure}
\begin{center}
\resizebox*{8cm}{!}{\includegraphics{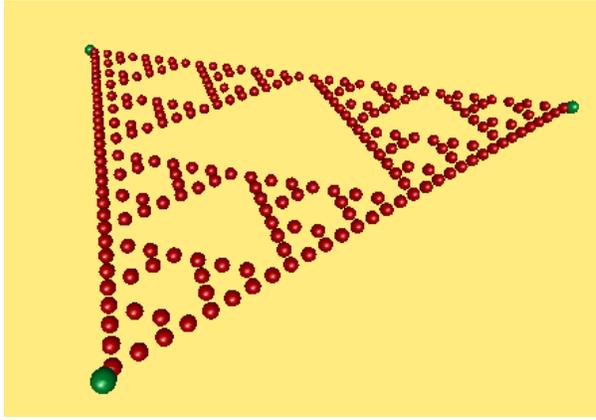}}
\caption{
To annihilate three defects (shown in green) at the corners of an equilateral 
triangle, one must flip the spins in a ``fractal'' membrane (containing sites 
shown in red) that stretches between the defects. For a triangle of size 
$2^\ell$, there are $3^\ell$ sites in the membrane. The annihilation of the 
three defects through quantum tunneling is a virtual process of order 
the number of sites that are involved (red sites). Hence, the 
amplitude for the quantum tunneling process vanishes exponentially with the 
``volume'' of the membrane. 
}
\label{fig:gasket}
\end{center}
\end{figure}

Through quantum tunneling, defect annihilation can occur via virtual 
processes in which the number of defects is strictly larger in the 
intermediate (virtual) steps. The order in perturbation theory in $g$ grows 
very fast with defect separation. An example is shown in 
Fig.~\ref{fig:gasket}; 
basically, to annihilate three defects at the 
corners of an equilateral triangle of size $\xi=2^\ell$, one must flip 
$3^\ell$ original spins laying on a mold defined by a Sierpinski gasket. 
(Notice that here the hierarchy is built staring from the microscopic
scale.) In perturbation theory the quantum recombination process has an
amplitude of order $(g/h)^{3^\ell}$, which leads to 
recombination/equilibration times 
\begin{equation}
t_{tun.} 
\sim
\exp\left[
  \ln\left(\frac{h}{g}\right) 
  \exp\left( \frac{\ln 3}{2\ln 2} \frac{h}{T} \right)
\right] 
\,,
\label{eq: tau q fragile}
\end{equation}
which grows extremely fast as the temperature is lowered. 
Again, we learn from this simple estimation that quantum tunneling is 
less effective than classical sequential processes in thermalizing the system. 

In a finite size system, the limiting time scale will be set by $\xi \sim L$, 
whereby 
\begin{equation}
\tau_q 
\sim 
\left(\frac{g}{h}\right)^{L^{\log_2(3)}}
= 
\exp\left[ L^{\log_2(3)} \ln\left(\frac{g}{h}\right) \right]
\end{equation}
leading to a quantum mechanical exponential dependence on system size in 
contrast with the polynomial scaling of the classical (activated) time 
scales that become available at any finite temperature,  
\begin{equation}
\tau_c 
\sim 
\exp\left[ \frac{h \log_2(L)}{T} \right] 
= 
L^{h / T \ln 2}
\,, 
\end{equation}
as in the long time scale limit $\xi \to L$, the classical energy barriers 
$\ell h$ tend to the value $h \log_2(L)$. 
%
%
%%%%%%%%%%%%%%%%%%%%%%%%%%%%%%%%%%%%%%%%%%%%%%%%%%%%%%%%%%%%%%%%%%%%%%%%%%%%%%%

\section{\label{sec: characterisation}
Note on perturbative approaches
        }
It is interesting to comment briefly on more generic approaches to study 
equilibration in quantum systems coupled to a bath, specifically in regards 
to quantum glasses with exponential time scales. 

Quantum systems in contact with a reservoir are characterized by 
mixed ensembles of states that are best described using von Neumann's density 
matrix formalism~\cite{vonNeumann1927,Blum_book}. 
At equilibrium, the density operator is given by 
$\hat{\rho} = e^{-\beta \hat{H}}/Z$ in the canonical ensemble, where 
$\hat{H}$ is the Hamiltonian of the system, $\beta = 1/T$ is the inverse 
temperature of the reservoir (and of the system of interest, 
\textit{at equilibrium}), and $Z = \tr(e^{-\beta \hat{H}})$ is 
the canonical partition function. 
The expectation value of any operator $\hat{\mathcal{O}}$ in this mixed 
ensemble is given by $\tr(\hat{\rho} \hat{\mathcal{O}})$. 
Therefore, the characterisation of the properties of a quantum system at 
equilibrium is essentially a spectral problem. 
Describing the low temperature properties of a system, for example, requires 
the understanding of the ground state, its symmetries (or lack thereof) and 
its quantum orders~\cite{Wen2002}, and its low lying excitations. 
In particular, at zero temperature, the system goes to its ground state. 
In the case of a quantum many-body system, if parameters in the Hamiltonian 
$\hat{H}$ are tuned, the ground state can change its symmetries or its 
quantum orders through quantum phase transitions. 

The time evolution of a system $s$ in contact with a reservoir $b$ and
total Hamiltonian
\beq
H = H_s + H_b + V \equiv H_0 + V
\eeq
is described by the Liouville-von Neumann equation 
\beq
\frac{d \rho(t)}{dt} = -i \left[ H , \rho(t) \right] 
, 
\label{eq: full Liouville}
\eeq
where $\rho(t) = \exp(i H t)\; \rho\; \exp(- i H t)$ is the time-dependent 
density matrix of the system plus bath, and we use the convention $\hbar = 1$. 
(Without loss of generality, we assume that $H$ does not depend directly 
on time.) 
It is customary to recast Eq.~\eqref{eq: full Liouville} in the interaction 
picture format, 
\bea
\rho_I(t) 
&\equiv& 
e^{-i H_0 t} \; e^{i H t} \; \rho \; e^{- i H t} \; e^{i H_0 t} 
\\ 
V_I(t) &\equiv& e^{-i H_0 t} \;V \; e^{i H_0 t} 
\eea
whereby 
\beq
\frac{d \rho_I(t)}{dt} = -i \left[ V_I(t) , \rho_I(t) \right] 
. 
\label{eq: full Liouville interact}
\eeq
Note that $\rho_I(0) = \rho(0) = \rho$. 
If we are interested in the evolution of the system alone, we need to trace 
over the degrees of freedom of the reservoir, 
\bea
\rho^{(s)}_I(t) &\equiv& \tr_b [\rho^{}_I(t)] 
\\ 
\frac{d \rho^{(s)}_I(t)}{dt} 
&=& 
-i \,\tr_b \left[ V_I(t) , \rho_I(t) \right] 
. 
\label{eq: sys Liouville}
\eea

The right hand side of Eq.~\eqref{eq: sys Liouville} results 
in a linear super-operator acting on $\rho^{(s)}_I(t)$. By comparison with 
its classical counterpart, i.e., the transition matrix in a 
Master equation, the eigenvalues of the super-operator determine the 
equilibration rates of the quantum system coupled to the bath. 
Following Ref.~\onlinecite{Castelnovo2010}, one could then symmetrise 
the super-operator and reinterpret it as a fictitious quantum `Hamiltonian' 
acting on the space of density matrices $\rho$. The appearance of diverging 
time scales in the quantum system would then correspond to a quantum phase 
transition in the associated `Hamiltonian'. 
Note that at the quantum level, the correspondence between the original 
Hamiltonian and the fictitious one occurs interestingly in the same number 
of dimensions for both systems. 

Unfortunately, taking the trace of the right hand side of 
Eq.~\eqref{eq: sys Liouville} to find the super-operator 
acting on $\rho^{(s)}_I(t)$ is in general a tall order. 
A common approach consists of expanding the commutator 
$[V_I(t) , \rho_I(t)]$ perturbatively in the interaction between system and 
reservoir, leading to the recursive equation 
\bea
\frac{d \rho_I(t)}{dt} 
&=& 
-i \left[ V_I(t) , \rho_I(0) \right] 
\nonumber \\
&+& 
\sum^{\infty}_{n=1} (-i)^{n+1} 
  \int^t_0 \!\! d\tau_1 
  \int^{\tau_1}_0 \!\! d\tau_2 \ldots \! 
  \int^{\tau_{n-1}}_0 \!\!\! d\tau_n \, 
    [V_I(t),[V_I(\tau_1), \ldots [V_I(\tau_n),\rho_I(0)]] \ldots ] 
\nonumber \\ 
&=& 
-i \left[ V_I(t) , \rho_I(0) \right] 
- 
\int^t_0 d\tau [V_I(t),[V_I(\tau),\rho_I(0)]] 
+ 
\ldots 
. 
\label{eq: full Liouville expanded}
. 
\eea
Carrying out the trace becomes now feasible under the conventional assumption 
that the initial density matrix factorises into system and bath, 
$\rho_I(0) = \rho^{(s)} \otimes \rho^{(b)}$. 
However, one needs to consider that each order in the expansion enables 
tunnelling processes in the system-bath interaction $V$ up to that very same 
order. As discussed in Sec.~\ref{sec: quantum glassiness} and as seen in the 
concrete examples in Sec.~\ref{sec: quantum glass examples}, the relaxation 
processes typical of quantum glasses require tunnelling under 
barriers whose width scales with system size. Such processes would be 
all together forbidden at any finite order in the perturbative expansion. 
One could view this as a \textit{signature of quantum glassiness}: 
the appearance of disconnected sectors in the quantum dynamics of the 
system at any finite order in perturbation theory. 
%
%
%%%%%%%%%%%%%%%%%%%%%%%%%%%%%%%%%%%%%%%%%%%%%%%%%%%%%%%%%%%%%%%%%%%%%%%%%%%%%%%

\section{\label{sec: conclusions}
Conclusions
        }
There are plenty of systems in nature that recalcitrantly avoid
equilibration with the environment. When their time scales become 
impractically large to measure, these systems are conventionally referred 
to as glasses. 

In order to investigate how this phenomenon comes about, it is reasonable 
to distinguish between different classes of slow dynamics according to the 
way that their equilibration times scale with their physical size. 
One can draw a parallel with computational complexity. 
When the system does not manage to equilibrate in ${\rm poly}(L)$ time, 
it is because nature's algorithm for (local) dynamics is not efficient. 
Systems with $\exp(L)$ time scales are computationally hard given nature's 
resources. 
[We note in passing that, just as in the case of the definition of 
computational complexity classes, one must allow for considerations of 
practical importance: time scales can diverge as a function of temperature 
while remaining polynomial in $L$.] 
%For quantum systems, however, the
%exponential {\it vs.} polynomial definition is well suited at $T=0$.

What happens when temperature is lowered all the way down to $T=0$? 
Because absolute zero temperature freezes out all classical
activation processes, quantum tunneling is all that is left to
dynamics. The relevant issue then becomes which kind of tunneling
barriers remain to surpass so as to reach the ground state in the
presence of a zero temperature bath. It is the nature of these 
barriers that separates systems with relaxation times that are
exponential vs. polynomial in $L$, and therefore separates 
systems that are quantum glasses (hard, for nature's algorithm) from those 
that are not. 

In this paper we discussed these issues and constructed explicit examples 
of systems with exponential time scales at zero temperature. 
These systems are devoid of disorder or local symmetry breaking, but rather 
exhibit a non-local form of order known as topological order. 
As such, they ought to be considered examples of topological quantum 
glasses. 
We showed how our {\it local} quantum Hamiltonians resist 
equilibration with a thermal bath and glassiness is ``protected'' against 
thermal equilibration with {\it any} bath, provided that it couples 
{\it locally} with the {\it physical} degrees of freedom of the system. 

It is often believed that quantum tunneling provides an escape route against
classical dynamical slowdown caused by thermal energy barriers as the
temperature is lowered. In the systems presented here, classical
sequential processes are more effective than quantum tunneling processes in
the limit $T \to 0$. The reason for the freezing of quantum tunneling is
that equilibration is through {\it activated} defect recombination; as 
the density of defects decreases at low temperatures, the barrier 
{\it widths} increase and eventually become as large as the size of the 
system.

We also discussed briefly generic approaches to study equilibration in 
quantum systems in contact with a reservoir, starting from the 
Liouville-von Neumann equation. Following this route one can in principle 
arrive at the quantum mechanical counterpart to the classical Master 
equation for the equilibration of probabilities. 
The role played by the transition matrix is here taken by a super-operator, 
whose spectrum controls the relaxation properties of the system. 
It is intriguing to speculate that a quantum-to-quantum mapping akin 
to the classical-to-quantum correspondence discussed in 
Ref.~\onlinecite{Castelnovo2010} would lead to the formulation of a 
fictitious quantum mechanical system in the same number of dimension of 
the original one, where quantum dynamical transitions would appear as 
static, zero temperature quantum phase transitions. 
However, constructing the super-operator is in general a tall order and 
at best one can do so order by order in perturbation theory. 
One should note however that truncating the expansion to any
fixed order in perturbation theory on the system/bath coupling disconnects 
the space of states. 
In order to study quantum glassy systems, 
it is necessary to go to all the way to order $L$ in the
perturbation expansion, accessing matrix elements that are
exponentially small, and therefore time scales that are exponentially
large in $L$. 

%\textbf{***CC: still missing:}
%Add references on finite $T$ quantum memories?

%\textbf{***CC: do we want to put out there the idea of protecting 
%topological quantum memories through glassiness? it may require a fair 
%bit of writing.} 

%
%
%%%%%%%%%%%%%%%%%%%%%%%%%%%%%%%%%%%%%%%%%%%%%%%%%%%%%%%%%%%%%%%%%%%%%%%%%%%%%%

\section*{
Acknowledgement(s)
         }
This work was supported in part by EPSRC Postdoctoral Research Fellowship 
EP/G049394/1 (C.C.), and by DOE Grant DEFG02-06ER46316 (C.C.). 
%
%
%%%%%%%%%%%%%%%%%%%%%%%%%%%%%%%%%%%%%%%%%%%%%%%%%%%%%%%%%%%%%%%%%%%%%%%%%%%%%%

%\section*{Note(s)}
%
%
%%%%%%%%%%%%%%%%%%%%%%%%%%%%%%%%%%%%%%%%%%%%%%%%%%%%%%%%%%%%%%%%%%%%%%%%%%%%%%

%\appendices

%\section{\label{app: ...}
%...
%        }

%
%
%%%%%%%%%%%%%%%%%%%%%%%%%%%%%%%%%%%%%%%%%%%%%%%%%%%%%%%%%%%%%%%%%%%%%%%%%%%%%%

%\cite[see][and references therein]{fzf88}

\bibliographystyle{tPHM}
\bibliography{quantglass}% Produces the bibliography via BibTeX.

\iffalse

\fi

%\label{lastpage}

\end{document}